# Black hole entropy, the black hole information paradox, and time travel paradoxes from a new perspective

Roland E. Allen

*Physics and Astronomy Department, Texas A&M University, College Station, TX USA*

allen@tamu.edu

# Black hole entropy, the black hole information paradox, and time travel paradoxes from a new perspective

Relatively simple but apparently novel ways are proposed for viewing three related subjects: black hole entropy, the black hole information paradox, and time travel paradoxes. (1) Gibbons and Hawking have completely explained the origin of the entropy of all black holes, including physical black holes – nonextremal and in 3-dimensional space – if one can identify their Euclidean path integral with a true thermodynamic partition function (ultimately based on microstates). An example is provided of a theory containing this feature. (2) There is unitary quantum evolution with no loss of information if the detection of Hawking radiation is regarded as a measurement process within the Everett interpretation of quantum mechanics. (3) The paradoxes of time travel evaporate when exposed to the light of quantum physics (again within the Everett interpretation), with quantum fields properly described by a path integral over a topologically nontrivial but smooth manifold.

## 1. Introduction

The three issues considered in this paper have all been thoroughly discussed by many of the best theorists for several decades, in hundreds of extremely erudite articles and a large number of books. Here we wish to add to the discussion with some relatively simple but apparently novel ideas that involve, first, the interpretation of the Euclidean path integral as a true thermodynamic partition function, and second, the Everett interpretation of quantum mechanics. We believe that some mysteries and paradoxes that appear to arise in conventional views of quantum fields can be resolved if one switches to the new perspective emphasized here, in the same way that early mysteries in astronomy (such as the apparently retrograde motion of Mars) were simply resolved by the Copernican interpretation.

**2. Black hole entropy**

The well-known attempts to explain the Bekenstein-Hawking entropy of black holes [1,2] (including those with angular momentum and charge) have usually involved extremely sophisticated arguments in string theory, loop quantum gravity, etc. -- and yet have still failed to explain the entropy of real black holes, which are not extremal and which dwell in in 3-dimensional space. These efforts are in dramatic contrast to the well-known simplicity of the formulas for the Bekenstein-Hawking entropy and Hawking temperature.

Less often emphasized is the brilliant demonstration of Gibbons and Hawking [3] that the Euclidean path integral $Z_{BH}$ of a general black hole yields exactly the right form for the Bekenstein-Hawking entropy, if $Z_{BH}$ can be interpreted as a true thermodynamic partition function (ultimately based on microstates).

We now show that this result follows from an earlier paper [4], which has other virtues including the prediction of a credible dark matter candidate [5,6], and whose equation (3.29) states

$$S_L = -S.  \qquad (1)$$

(In this section, natural units are used, with $\hbar = c = k = 1$.) We have dropped a constant which has no effect on physical properties, and have changed the notation slightly to emphasize a central feature of Ref. 4: Although the initial *path integral* has a *Euclidean* form, involving a Boltzmann entropy $S$, the resulting *action* $S_L$ automatically yields the proper *Lorentzian* form for the action of the physical fields that are derived in the remainder of the paper.

The Lorentzian path integral for the gravitational field has the form

$$Z_L = \int \mathcal{D}g \, e^{iS_L} \quad , \quad S_L = \int d^4x \, \mathcal{L} \quad , \quad \mathcal{L} = \mathcal{T} - \mathcal{V}$$

where the Lagrangian density $\mathcal{L}$ consists of terms $\mathcal{T}$ which contain time derivatives and contributions $\mathcal{V}$ which do not. (We could also write this in the form familiar from elementary mechanics, $S_L = \int dt \, L, L = T - V$. The conclusions below also hold with the Gibbons–Hawking–York boundary term included.) For a stationary system, with $\mathcal{T} = 0$, the usual transformation to a Euclidean path integral (a Wick rotation in time with $t \to -it$ and $x^0 = t$) implies

$$\int dx^0 d^3x \to -i \int dx^0 d^3x$$

$$Z_E = \int \mathcal{D}g \, e^{-S_E} \quad , \quad S_E = \int d^4x \, \mathcal{L}_E \quad , \quad \mathcal{L}_E = \mathcal{V} \quad \text{for a stationary system.}$$

The initial path integral of Ref. 4, on the other hand, has the above Euclidean *form*, but with the quantity $S$ of (1) rather than $S_E$.

In Ref. 4 it is shown that the interpretation of (1) leads, through a set of nontrivial steps, to standard physics in a *Lorentzian* picture. Here, however, we are concerned with the connection to the argument of Ref. 3, which is based on a *Euclidean* action. The main point is that $\mathcal{T} = 0$ for a system whose time derivatives are zero, so that generically

$$S_E = -S_L = S \quad \text{for stationary systems.} \tag{2}$$

including Kerr-Newman black holes with angular momentum and charge.

$S$ is derived as the entropy in a "microcanonical" picture, but the argument of Ref. 3 is in a "canonical" picture, for a black hole with a given mass, charge, and angular momentum. These (conserved extensive) quantities can be treated in the same way as, e.g., energy, number of particles, etc. are treated in the standard approach employing the canonical or grand canonical ensemble. (The ensemble itself has a Boltzmann entropy, which is maximized. The intensive quantities – here temperature, potential, and angular velocity – are first treated as Lagrange multipliers, and then identified as physical quantities through a comparison with the first law of thermodynamics, in which

$$dE = dq - dw \quad , \quad dq = TdS_{BH} \quad , \quad -dw = \Omega dJ + \Phi dQ$$

in a standard notation.) In this context, the Euclidean action for an individual system is related to a free energy $W$, and the new feature from Ref. 4 in this context is that the stationary systems in an ensemble (which are actually field configurations) have well-defined microstates.

Here we do not repeat the arguments of Ref. 3, since they are quite nontrivial and this paper is readily available. The final result is its Eq. (3.13) for the entropy (under the assumption that $W$ can somehow be interpreted as the true free energy based on microstates), which exactly agrees with the expected Bekenstein-Hawking entropy:

$$S_{BH} = \frac{1}{4}\frac{A}{\ell_P^2}$$

where $\ell_P$ is the Planck length. One can also see directly from (3.7)-(3.11) of Ref. 3 that

$$S_{BH} = W/T = S_E$$

for a nonrotating (and possibly charged) black hole, just as in (2). (For a rotating black hole, a "quasi-Euclidean" action is evaluated in a rotating frame of reference.) The canonical approach of Ref. 3, however, yields both the Bekenstein-Hawking entropy and the Hawking temperature in their expected forms.

The conclusion, then, is that Gibbons and Hawking have completely explained the entropy of a general black hole for any theory, like that of Ref. 4, where the Euclidean action and path integral are ultimately obtained from a Boltzmann entropy based on microstates.

**2. Black hole information paradox**

This issue has also received an enormous amount of attention and been subjected to many sophisticated interpretations [7,8]. If one considers the Hawking radiation as consisting of particles detected at the position of a distant observer, the distribution of these particles is thermal, with a temperature that is consistent with the Bekenstein-Hawking entropy considered above. But the detection of these particles is just like the detection of particles emitted by other quantum systems, from radioactive nuclei (which determine the fate of Schrödinger's cat) to excited atoms undergoing spontaneous emission to the quantum fluctuations in the very early universe which are thought to be responsible for the large scale structures now spread across the sky.

In each case, one has the "paradox" of wave-particle duality and the measurement problem, which has been the subject of a vast number of papers and many books during the past century, starting with the very early misgivings of Einstein. As is argued in the next section, there are only two logically consistent ways to resolve this "paradox": (i) some drastically new physical picture of reality or (ii) the Everett interpretation, in which standard quantum theory is accepted as fully correct, with no magical wavefunction collapse when the state of a quantum system is observed.

In the full quantum picture of the Everett interpretation, before observation by a distant observer, Hawking radiation is described by amplitudes of quantum fields which evolve deterministically, with no loss of unitarity or information.

There is as yet no generally accepted theory of quantum gravity, so there is no way to follow in detail the evolution of, e.g., a proton-sized black hole until the emission of Hawking radiation eventually results in its complete evaporation or explosion. It is conjectured that intense processes (near points that are singularities in a classical description) will result in baryon number nonconservation etc., as is natural in a grand-unified theory. But there is no reason to believe that these processes will violate the unitarity of any normal quantum theory, with the time dependence governed by a Hamiltonian or path integral. This is even true in very exotic scenarios like the birth of baby universes.

To a distant observer who detects the Hawking thermal emission of particles, it will seem that an original pure quantum state evolves into a mixed state described by a density matrix. But in the complete Everett quantum picture, the full system remains in a pure state – even though it is a very complicated one that is counter to ordinary human intuition. The behavior of the black hole and the matter in it is extremely exotic, and should probe regions of physics that are not yet understood in detail. But there is no reason to believe that the most fundamental aspects of quantum theory will not prevail and dictate deterministic evolution of a pure state with no loss of information.

**2. Time travel paradoxes**

In this section we relax into a more popular mode because the subject overlaps strongly with popular culture. The well-known treatments of time travel in science fiction are all based on a classical mindset, with a single preferred trajectory of the world through spacetime. Even when a branching between alternative realities is shown, as in the blackboard presentation of "Back to the Future, Part II", only one is understood to be the true course that the world follows. As stated in a summary [9], "Marty and Doc must return to 1955 to keep an alternate version of 1985 from forming." The Terminator series seems to have the same philosophy.

Most serious discussions of time travel have also been based on a picture with only one classical reality, which must somehow be made consistent to avoid paradoxes, as in Robert Heinlein's "By His Bootstraps" and "All You Zombies". The extremely clever and amusing contrivances required to achieve consistency in these stories are, to put it mildly, unlikely to be realized through natural processes. And, more importantly, experiment has now disconfirmed classical physics and demonstrated that we live in a quantum universe [10-19].

Our current view of nature is that it is composed entirely of quantum fields (although the proper description of quantum gravity is still undecided). All physics can be regarded as ultimately derived from an enormous path integral with the form

$$Z = \int \mathcal{D}\Phi e^{iS[\Phi]/\hbar} \quad (3)$$

where the action $S$ is now a functional of all fields $\Phi$ over all spacetime. The gravitational field is described by a metric tensor $g_{\mu\nu}$. The other fields will be represented by $\phi$. $Z$ contains all physically distinct field configurations that are consistent with whatever boundary conditions are imposed.

For a toy model, consider a universe with only two coordinates – one for space and one for time. There are two independent aspects to its geometry – first the manifold, or set of points with a specific topology, and second the metric, which measures the intervals between points. We could have nontrivial topologies such as a sphere, but for simplicity let us first focus on a flat plane.

There is actually, and importantly, a third aspect to the geometry of the universe – the geometry of the matter and force fields. In our toy model let us have only a single real field $\phi$, represented by a line of real numbers – or fiber – which is perpendicular at each point to the horizontal *xt* plane of the preceding paragraph, and thus parallel to a vertical $\phi$ axis. The classical equation of motion for this field is determined by extremalizing the action $S$ (and for the purpose of illustration we can invent an action to make this equation nontrivial). The classical trajectory of the field $\phi$ is then a well-defined function $\phi(x,t)$.

But in quantum mechanics the field $\phi$ follows *all* trajectories – i.e, all paths in the path integral, with each weighted by the factor $e^{iS/\hbar}$. (This is equivalent to saying that $\phi$ ranges over all possible values at every point in the $x,t$ plane.) As Feynman pointed out, the classical path is typically dominant for $S \gg \hbar$, because it exhibits constructive interference, with strong destructive interference along the paths that deviate significantly from it. So our everyday experience in the macroscopic world of ordinary life leads to the expectation that matter and force fields will follow only a single path through spacetime, determined by the laws of classical physics.

There are, however, cases when the multiple paths required by quantum physics are followed even by macroscopic objects. One example is the measurement process, which by definition (in the present context) means that the state $|M\rangle$ of a macroscopic system (like human observer plus Schrödinger's cat) is entangled with the state $|m\rangle$ of a microscopic system (like an atomic nucleus), so that the total state has the form

$$(|M\uparrow\rangle|m\uparrow\rangle + |M\downarrow\rangle|m\downarrow\rangle)/\sqrt{2} .$$

After a measurement, there are then two diverging branches in the path integral – the $\uparrow$ branch and the $\downarrow$ branch – e.g. following the versions of an observer which see a live or a dead cat.

The recognition that quantum physics applies to all physical systems, including human beings, is called the Everett interpretation [20-26], and it has acquired increasing acceptance (among those who have thought deeply about the issues) as the decades pass and it becomes increasingly clear that quantum physics does describe nature in its entirety.

Despite the vast number of supposedly distinct proposals that have been put forth, simple (clear-headed) logic ultimately dictates that either (i) the Everett interpretation is correct or (ii) there is new physics that will drastically change our worldview (by, e.g., somehow driving collapse of the quantum wavefunction during a measurement). But there is not even the slightest hint of such physics in the extremely large number of exquisitely precise experiments testing many different aspects of quantum physics.

There have, of course, been many discussions of the quantum mechanics of time travel [27,28], but usually with an emphasis on the connection to quantum computation or to obtaining self-consistent solutions with closed timelike loops. (Even a very modest review of the literature on the interpretation of quantum mechanics or the physics of time travel would be several orders of magnitude longer than permitted in the present contribution.) Let us consider this topic in the simplest and most direct way possible, via standard quantum field theory and the path integral of Eq. (3).

We postulate the wormhole mechanism for time travel invented by Kip Thorne [29-31], who is shown in Fig. 1. (The argument given below is a more detailed version of one of the multiple possibilities raised by David Deutsch in Ref. 27 and stated with more commitment by him elsewhere, but given here with an emphasis on path integrals on a smooth manifold with a nontrivial topology. Of course, the idea that time travel results in parallel universes – but without any physical justification – goes back in the science fiction literature to at least the early 1950s, before the work of Everett.) We assume that a Thorne wormhole can be used to achieve time travel and that it is traversable – i.e., somehow stable and large enough to accommodate classical field configurations like human observers.

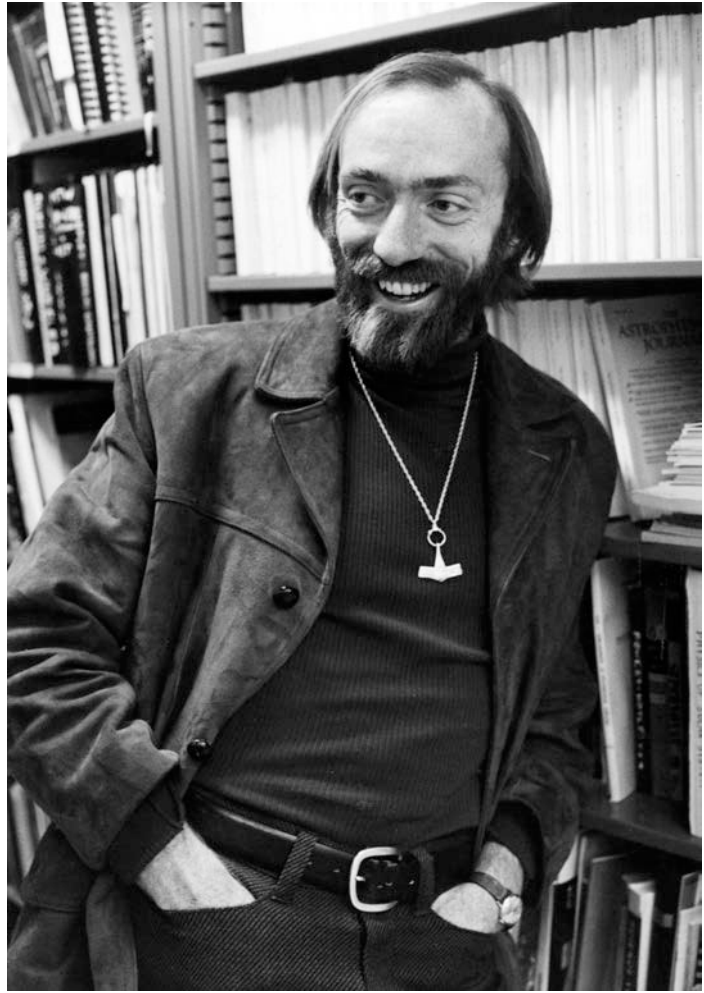

Figure 1. Kip Thorne in his office at the California Institute of Technology. Courtesy of the Archives, California Institute of Technology.

Here we consider path integrals with quantum fields (fiber bundles) defined over a manifold qualitatively like that of Fig. 2. At every point of the 4-dimensional manifold all of the fields are well-defined and vary smoothly. An object traversing the wormhole will then pass smoothly from one opening to the other. We therefore disagree with the proposal that an observation will necessarily send different parts of, e.g., a human body into different Everett branches [32], ripping it asunder, unless there is some physical mechanism analogous to tidal forces that divides the different parts of the body. There is not a sudden either-or transition from one time to another, but instead the smooth transition along a nearly classical path.

Figure 2 schematically represents a topology of the (1+1) dimensional toy model which permits time travel – with a wormhole which can be entered at time $t_2$ and exited at an earlier time $t_1$. (A third dimension is now needed to depict the nontrivial two-dimensional manifold. The wormhole shown here is adequate for a qualitative argument, but is not a realistic version like the kind invented by Kip Thorne.) A human observer Alice who performs this feat in our real (3+1) dimensional world can be interpreted as a configuration $\phi_{\text{Alice}}$ of quantum fields undergoing a transition from $\phi_{\text{Alice}}(t_2)$ to $\phi'_{\text{Alice}}(t_1)$.

Suppose that Alice travels one day into the past, to a time when the world was in state $\Phi_{\text{world}}(t_1)$ and she was in state $\phi_{\text{Alice}}(t_1)$. Before she exerts any influence, all will be the same as she remembers.

But suppose she then encounters her previous self. After a time $\Delta t$, her previous self will be in a state $\phi''_{\text{Alice}}(t_1 + \Delta t)$ (on a diverging quantum path) which is different from the state $\phi_{\text{Alice}}(t_1 + \Delta t)$ that she remembers having experienced. And after one day has passed, the whole world will be in a state $\Phi''_{\text{world}}(t_1 + 1 \text{ day})$ rather than its state $\Phi_{\text{world}}(t_1 + 1 \text{ day})$ when she departed to the past. She and her younger self will be in states $\phi'_{\text{Alice}}(t_1 + 1 \text{ day})$ and $\phi''_{\text{Alice}}(t_1 + 1 \text{ day})$ respectively.

The fields of the world that she now observes will be on a different quantum branch in the full path integral (2). There are no paradoxes, even if Alice were now to travel back 60 years and kill her grandmother before her mother was born (in which case Alice will only exist on the original branch and not the altered one).

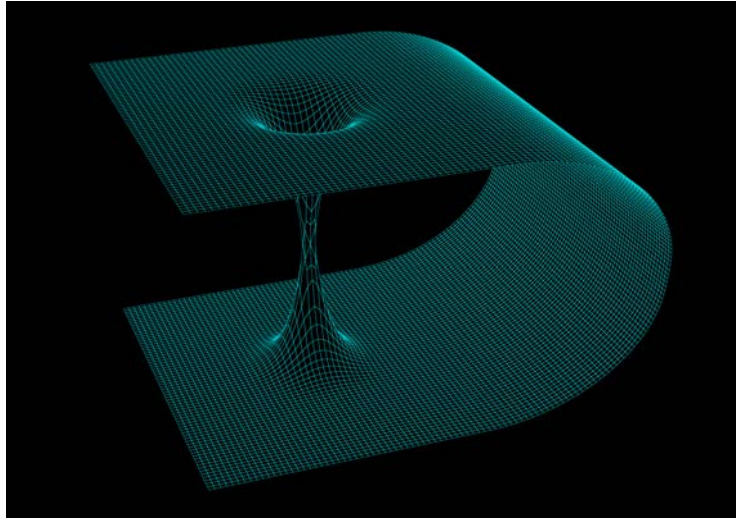

Figure 2. Purely schematic representation of a wormhole which connects two different times near the same point in space, permitting travel from the later to the earlier time, in the toy model of a (1+1) dimensional universe. Credit: Shuttercock.

There is no need for a consistent classical history, and (in the present context) such a history has vanishingly low probability. If Alice wishes to retrace her path into her original branch (i.e. the universe whose history is familiar to her), it will ordinarily be difficult to find the precise path that accomplishes this. In general, every traversal of the wormhole will end on a different branch, perturbed by the presence of $\phi'_{\text{Alice}}(t)$.

Stephen Hawking proposed a chronology protection conjecture, saying that he wanted to make the universe safe for historians. But the current picture is vastly better in this respect, since it provides unlimited opportunities and employment for historians, with a potentially infinite set of quite distinct histories.

The difference between this true quantum picture and the fictional one of "Back to the Future, Part II" is that all branches are equally real. The quantum picture seems to be more nearly consistent with the basic plot of "Déjà Vu", a movie that leaves one with the disturbing recognition that there is an alternate reality in which both of the principal characters have suffered horrific deaths.

Our principal conclusion in this section: The paradoxes of time travel evaporate when exposed to the light of quantum physics, with quantum fields properly described by a path integral over a topologically nontrivial but smooth manifold.

But there are still two major questions concerning time travel. The first is whether it will be possible even for our most remote descendants. To create a wormhole large enough to accommodate an observer would apparently require very new physics – either a new formulation of gravity or else Kip Thorne's "exotic matter" with negative energy [29-31].

The second question is why our descendants might wish to travel through time. But this is analogous to asking why we wish to explore space. Like our ancestors, we are motivated by both curiosity and the desire to push outward to new frontiers. So time travel may, in the extremely remote future, turn out to be another natural part of the human story – the ultimate version of exploration, perhaps combined with paleontology (if early wormhole mouths could be found or created), if not tourism.

Acknowledgement: This work was largely stimulated by a workshop on black hole entropy and quantum thermodynamics sponsored by the Institute for Quantum Science and Engineering (IQSE) at Texas A&M University, and by related sessions at the 48[th] Winter Colloquium on the Physics of Quantum Electronics (PQE-2018), featuring Don Page, Gerald Moore, Marlan Scully, and Stephen Fulling.

---

[1] Don N. Page, "Hawking Radiation and Black Hole Thermodynamics", New J. Phys. 7, 203 (2005).

[2] Marlan O. Scully, Stephen Fulling, David Lee, Don Page, Wolfgang Schleich, and Anatoly Svidzinsky, "Radiation from Atoms Falling into a Black Hole", arXiv:1709.00481 [quant-ph].

[3] G.W. Gibbons and S. W. Hawking, "Action integrals and partition functions in quantum gravity", Phys. Rev. 15, 2752 (1977).

[4] Roland E. Allen, "Predictions of a fundamental statistical picture", arXiv:1101.0586 [hep-th].

[5] R. E. Allen and A. Saha, "Dark matter candidate with well-defined mass and couplings", Mod. Phys. Lett. A 32, 1730022 (2017).

[6] Roland E. Allen, "Saving supersymmetry and dark matter WIMPs -- a new kind of dark matter candidate with well-defined mass and couplings", Phys. Scr. (in press).

[7] William G. Unruh and Robert M. Wald, "Information loss", Rep. Prog. Phys. 80, 092002 (2017).

[8] S. W. Hawking, M.J. Perry, and A. Strominger, "Soft Hair on Black Holes", Phys. Rev. Lett. 116, 231301 (2016).

[9] http://backtothefuture.wikia.com/wiki/Back_to_the_Future_Part_II.

[10] S. J. Freedman and J. F. Clauser, "Experimental Test of Local Hidden-Variable Theories", Phys. Rev. Lett. 28, 938-941 (1972).

[11] E. S. Fry and R. C. Thompson, "Experimental Test of Local Hidden-Variable Theories", Phys. Rev. Lett. 37, 465-468 (1976).

[12] A. Aspect, P. Grangier, and G. Roger, "Experimental Realization of Einstein-Podolsky-Rosen-Bohm Gedankenexperiment: A New Violation of Bell's Inequalities", Phys. Rev. Lett. 49, 91–94 (1982).

[13] B. Hensen, H. Bernien, A. E. Dréau, A. Reiserer, N. Kalb, M.S. Blok, J. Ruitenberg, R. F. L. Vermeulen, R. N. Schouten, C. Abellán, W. Amaya, V. Pruneri, M. W. Mitchell, M.

Markham, D. J. Twitchen, D. Elkouss, S. Wehner, T. H. Taminiau, and R. Hanson, "Loophole-free Bell inequality violation using electron spins separated by 1.3 kilometres", Nature 526, 682–686 (2015).

[14] M. Giustina, M. A. M. Versteegh, S. Wengerowsky, J. Handsteiner, A. Hochrainer, K. Phelan, F. Steinlechner, J. Kofler, J-Å. Larsson, C. Abellán, W. Amaya, V. Pruneri, M. W. Mitchell, J. Beyer, T. Gerrits, A. E. Lita, L. K. Shalm, S. W. Nam, T. Scheidl, R. Ursin, B. Wittmann, and A. Zeilinger, "Significant-Loophole-Free Test of Bell's Theorem with Entangled Photons", Phys. Rev. Lett. 115, 250401 (2015).

[15] L. K. Shalm, E. Meyer-Scott, B. G. Christensen, P. Bierhorst, M. A. Wayne, M. J. Stevens, T. Gerrits, S. Glancy, D. R. Hamel, M. S. Allman, K. J. Coakley, S. D. Dyer, C. Hodge, A. E. Lita, V. B. Verma, C. Lambrocco, E. Tortorici, A. L. Migdall, Y. Zhang, D. R. Kumor, W. H. Farr, F. Marsili, M. D. Shaw, J. A. Stern, C. Abellán, W. Amaya, V. Pruneri, T. Jennewein, M.W. Mitchell, P. G. Kwiat, J. C. Bienfang, R. P. Mirin, E. Knill, and S. W. Nam, "Strong Loophole-Free Test of Local Realism, Phys. Rev. Lett. 115, 250402 (2015).

[16] W. Rosenfeld, D. Burchardt, R. Garthoff, K. Redeker, N. Ortegel, M. Rau, and H. Weinfurter, "Event-Ready Bell Test Using Entangled Atoms Simultaneously Closing Detection and Locality Loopholes", Phys. Rev. Lett. 119, 010402 (2017).

[17] The BIG Bell Test Collaboration, C. Abelian et al., "Challenging local realism with human choices", Nature 557, 212 (2018).

[18] Louisa Gilder, *The Age of Entanglement: When Quantum Physics Was Reborn* (Vintage; 2009).

[19] Anton Zeilinger, *Dance of the Photons: From Einstein to Quantum Teleportation* (Farrar, Straus and Giroux; 2010).

[20] Hugh Everett, III, "'Relative State' Formulation of Quantum Mechanics", Rev. Mod. Phys. 29, 454-462 (1957).

[21] John A. Wheeler, "Assessment of Everett's 'Relative State' Formulation of Quantum Theory", Rev. Mod. Phys. 29, 463-464 (1957).

[22] Roland E. Allen, "Remarks on the Everett-Wheeler Interpretation of Quantum Mechanics," Am. J. Phys. 39, 842 (1971). Immediately after publication of this paper, John Wheeler sent the author a postcard asking that Everett be given exclusive credit.

[23] *The Many-Worlds Interpretation of Quantum Mechanics*, edited by B. S. DeWitt and R. Neill Graham (Princeton University Press, 1973).

[24] David Deutsch, *The Fabric of Reality: The Science of Parallel Universes--and Its Implications* (Penguin Books, 1998).

[25] *The Everett Interpretation of Quantum Mechanics: Collected Works 1955-1980 with Commentary*, edited by Jeffrey A. Barrett and Peter Byrne (Princeton University Press, 2012).

[26] *Many Worlds? Everett, Quantum Theory, and Reality*, edited by Simon Saunders, Jonathan Barrett, Adrian Kent, and David Wallace (Oxford, 2010).

[27] David Deutsch, "Quantum mechanics near closed timelike lines", Phys. Rev. D 44, 3197–3217 (1991).

[28] https://en.wikipedia.org/wiki/Quantum_mechanics_of_time_travel.

[29] Kip Thorne, *Black Holes and Time Warps: Einstein's Outrageous Legacy* (Norton,1995). The author has made the discussion of macroscopic (traversable) wormholes respectable through a series of technical papers -- and by employing them in the movie "Interstellar" and introducing them into modern science fiction, via a communication to Carl Sagan in 1985.

[30] Kip Thorne, *The Science of Interstellar* (Norton, 2014).

[31] *Wormholes, Warp Drives and Energy Conditions*, edited by Francisco S. N. Lobo (Springer, 2017).

[32] Allen Everett, "Time travel paradoxes, path integrals, and the many worlds interpretation of quantum mechanics", Phys. Rev. D 69, 124023 (2004).